\documentclass[showpacs,pra,twocolumn]{revtex4}
\usepackage{epsfig,amsmath}
\usepackage{subfigure}
\usepackage{graphicx}
\usepackage{dcolumn}
\usepackage{stmaryrd}
\usepackage{mathrsfs}
\usepackage{pifont}
\usepackage{amsthm}
\usepackage{amssymb}
\usepackage{bm}
\usepackage{latexsym}
\usepackage{hyperref}

\newcommand{\beq}{\begin{equation}}
\newcommand{\eeq}{\end{equation}}
\newcommand{\beqa}{\begin{eqnarray}}
\newcommand{\eeqa}{\end{eqnarray}}

\begin{document}

\title{Almost exact state transfer in a spin chain via pulse control}
\author{Zhao-Ming Wang$^{1,2}$, Marcelo S. Sarandy$^{3}$, Lian-Ao Wu$^{2,4}$\footnote{lianao.wu@ehu.es}}
\affiliation{$^{1}$ Department of Physics, Ocean University of China, Qingdao 266100,
China \\
$^{2}$ Department of Theoretical Physics and History of
Science, University of the Basque Country UPV/EHU, 48008, Spain \\
$^{3}$ Instituto de F\'{\i}sica, Universidade Federal Fluminense, Campus da Praia Vermelha, 24210-346, Niteroi, RJ, Brazil \\
$^{4}$ IKERBASQUE, Basque Foundation for Science, 48011
Bilbao, Spain}
\date{\today }

\begin{abstract}
Quantum communication through spin chains has been extensively investigated.
In this scenario, state transfer through linearly arranged spins connected by uniform nearest-neighbor couplings qualifies as a natural choice, with minimal control requirements. However, quantum states usually cannot be perfectly
transferred through a uniformly coupled chain due to the dispersion of the chain. Here,
we propose an effective quantum control technique to realize almost
exact state transfer (AEST) in a quantum spin chain. The strategy is to add a
leakage elimination operator (LEO) Hamiltonian to the evolution, which implements a
sequence of pulse control acting on a perfect state transfer subspace. By
using the one-component Feshbach PQ partitioning technique, we obtain the conditions
over the required pulses. AEST through a spin chain can then be
obtained under a suitable pulse intensity and duration.
\end{abstract}

\maketitle

%%%%%%%%%%%%%%%%%%%%

\section{Introduction}

Accurate control of a quantum system is crucial for
performing quantum tasks~\cite{Press,Domenico}, such
as adiabatic quantum computing~\cite{Farhi:01,Albash}, quantum thermodynamic processes~\cite{Kieu:04,Hu:19},
stable energy transfer in quantum batteries~\cite{Santos:19,Santos:19-2,Rosa:19},
speedup of quantum circuits~\cite{Nakahara,Santos:15}, acceleration of chemical reactions~\cite{Prezhdo}, among others.
In general, quantum control techniques are employed to constrain
a quantum evolution towards a target state or a target path~\cite{Wunotes}. For
example, in adiabatic quantum computation, the system is slowly evolved to be
maintained, with high probability, at the lowest-energy instantaneous eigenstate.
In order to speed up the adiabatic dynamics, shortcuts to adiabaticity can be
implemented through counter-diabatic control fields~\cite{Demirplak:03,Demirplak:05,Berry:09,Torrontegui:13}.
Counter-diabatic transitionless driving has been extensively realized.
In particular, it has been applied to speed up stimulated Raman
adiabatic passage in a solid-state lambda system~\cite{zhou} and in cold atoms~\cite{Du}.
More recently, counter-diabatic fields have also been used to
yield energy-optimal shortcuts to adiabaticity in trapped ions~\cite{Hu:18} and to
accelerate quantum gates in nuclear magnetic resonance~\cite{Santos:20}.

In the context of open system control, adiabatic speedup has recently been implemented
in a non-Markovian evolution through
an effective set of pulses on the system~\cite{Wang20182}.
In this scenario, a relevant set of techniques is provided by dynamical decoupling control,
such as Bang-Bang (B-B)~\cite{Viola,Vitali} and LEO control~\cite{Wu02}.
B-b control requires unbounded fast and strong pulses while LEOs are
introduced to counteract leakage from a subsystem to the rest of a
multilevel Hilbert space~\cite{Wu02,Byrd05,Campo13,Campo11}. In this case,
only finite pulse intensity and time interval are required, since the
effectiveness of the control only depends on the integral of the pulse
sequence in the time domain~\cite{Jing2014}. By adding a LEO Hamiltonian to
the adiabatic frame, the transitions between different instantaneous
eigenstates are restrained~\cite{Wang2018} and the system undergoes the adiabatic
passage. For a two-level system, the LEO Hamiltonian in the adiabatic frame is
equivalent to adding a global pulse on the Hamiltonian
in the experimental frame~\cite{Wang2018}.
This can be extended to the open system realm, where inverse engineering control
can be applied and a time-dependent Hamiltonian can be derived in order to guide the system to
attain an arbitrary target state at a predefined time~\cite{Wu2013}.

Quantum information processing tasks often require the active transmission
of a quantum state over relatively short distances~\cite{Kimble}. Spin chains
can serve as a suitable communication channel in these cases. In fact, reliable
quantum state transfer through a spin chain has been extensively studied
\cite{Bose,Wu20091,Yung,Wu20092,Wang2009}. It was Bose who first suggested
the use of an unmodulated spin chain as a channel for quantum information
transfer~\cite{Bose}. However, in most situations, this method does not allow for a
perfect state transfer. The fidelity will decrease quickly with increasing
length of the chain. Thus, a number of schemes have been proposed to improve the
transmission fidelity, such as by using a single local on-off switch actuator
\cite{Schirmer}, a suitable sequence of two-qubit gates at the
end of the chain~\cite{Burgarth}, optimal control to an external
parabolic magnetic field~\cite{Murphy}, and optimal state encoding~\cite%
{Wang2009}. More recently, the fastest transmission speed while maintaining high fidelities
has been achieved by reinforcement learning in a spin chain~\cite{Yun2018}.

The experimental platforms to realize spin systems include trapped ions
~\cite{Zeiher}, ultracold atoms in optical lattices~\cite{Grusdt}, quantum
dots~\cite{Veldhorst}, etc. For some physical systems, such as ultracold
atoms in optical lattices, the required couplings between different sites
can be created~\cite{Wang2013}. Engineering special couplings between
neighbor sites can realize perfect state transfer (PST)~\cite{Christandl} or
near PST~\cite{Wojcik,Oh,Wang2013}. Then, a PST or near PST trajectory can be
obtained. For some other systems, it might be inconvenient to
create the designed couplings. Here, we aim at creating a LEO Hamiltonian
by using either PST or near PST trajectories to realize almost exact state transfer (AEST),
being able to drive quantum communication through a set of pulses over a uniform spin chain.
By using the Feshbach PQ partition technique~\cite{Feshbach},
we obtain the conditions to be obeyed by the required pulses, which are provided
by the pulse intensity, pulse interval and total evolution time.
Throughout this analysis, we will consider various pulse patterns,
including rectangular pulses and sine function. Especially, we use the so-called
zero-energy-change pulse, in which the total integration is zero in one
period. In addition, we show that the pulse control scheme used in
Refs~\cite{Wang20182,Wang2018} to speed up adiabatic processes
can be generalized to the quantum communication domain.
\section{Construction of the LEO Hamiltonian}
%\section{Construction of the LEO Hamiltonian.}

Define a complete orthonormal
basis $\left\vert \Psi_{n}(0) \right\rangle$, $\langle \Psi_{m}(0) |
\Psi_{n}(0) \rangle =\delta _{mn}$. Suppose there exists a complete orthonormal
basis $\left\vert \Psi_{n}(t) \right\rangle$, $\langle \Psi_{m} (t)|
\Psi_{n}(t) \rangle=\delta _{mn}(t)$, and a one-to-one correspondence
between the states $|\Psi_{n} (0)\rangle$ and $| \Psi_{n}(t) \rangle$. For
example, $\left\vert \Psi_{n}(t) \right\rangle$ could be the instantaneous
eigenstates of a time-dependent Hamiltonian $H(t)$. Our target problem is to drive
the system along the $\left\vert \Psi_{1}(t) \right\rangle$ passage within a finite
time with high probability.

The LEO Hamiltonian can be constructed as
\begin{equation}
H_{LEO}(t)=c(t)\left\vert \Psi_{1}(t)\right\rangle \left\langle
\Psi_{1}(t)\right\vert ,
\end{equation}
where $c(t)$ is the external control function that describes a sequence of
control pulses. LEOs can be used to reduce errors from an encoded subspace $%
\left\vert \Psi_{1}(t)\right\rangle \left\langle \Psi_{1}(t)\right\vert $ to
the rest of the system's subspace, whether the pulses are ideal~\cite%
{Feshbach} or non-ideal~\cite{Jing2015PRL,Licheng}.
The total Hamiltonian is then
\begin{equation}
H(t)=H_{0}+H_{LEO}(t),
\end{equation}
where $H_{0}$ is the original Hamiltonian of the system, e.g., the local Hamiltonian of a
uniform spin chain.

\section{One-component dynamical equation}

Given a time-dependent Hamiltonian $H(t)$, the system dynamics is governed by its corresponding Schr\"{o}dinger equation
\begin{equation}
i | \dot{\Psi} (t) \rangle =H(t)\left\vert
\Psi(t)\right\rangle ,  \label{Eq3}
\end{equation}
with the dot symbol denoting time derivative and $\hbar =1$ throughout the paper.
The state vector $\left\vert \Psi (t)\right\rangle $ can be expanded in terms of the
time-dependent basis $\{\left\vert \Psi_{n}(t) \right\rangle\}$, which reads
\begin{equation}
\left\vert \Psi (t)\right\rangle =\sum\nolimits_{n}a_{n}(t)\left\vert \Psi
_{n}(t)\right\rangle ,  \label{Eq4}
\end{equation}
where $a_n(t)$ denotes a complex amplitude probability.
Substituting Eq.~(\ref{Eq4}) in Eq.~(\ref{Eq3}),
we obtain
\begin{equation}
i\dot{a}_{n}=\sum\limits_{m}[ \langle \Psi _{n} |
H(t) | \Psi _{m} \rangle - i \langle \Psi _{n}
| \dot{\Psi}_{m}\rangle] a_{m} .  \label{Eq5}
\end{equation}
Eq.~(\ref{Eq5}) can be rearranged in a vector form, with the left-hand-side written as
$\left\vert \Psi (t)\right\rangle=[a_{1},a_{2},...]^{^{\prime }}$ and the right-hand-side given
in terms of the effective Hamiltonian
$H_{n,m}(t)=\langle \Psi _{n} | H(t) | \Psi
_{m}\rangle -i\langle \Psi _{n} {| \dot{\Psi}
_{m}\rangle }$. To trace the footprint of $\left\vert \Psi
_{1}(t)\right\rangle $, one needs to find an exact one-component dynamical
equation for $a_{1}$. Feshbach P-Q partitioning technique provides an
effective approach to solve this problem~(see, e.g., Ref.~\cite{Jing:16}).

Using the $PQ$ partitioning, the $n$-dimensional state vector $\left\vert
\Psi (t) \right\rangle $ can be divided into two parts: a one-dimensional vector
of interest, $P(t)$, and the rest, an $(n-1)$-dimensional vector $Q(t)$.
The Hamiltonian can be split into three contributions $H(t)=H_{P}(t)+H_{Q}(t)+H_{L}(t)$, with
$H_{P}(t)$ and $H_{Q}(t)$ acting on the subspaces defined by $P(t)$ and $Q(t)$, respectively, and
$H_{L}(t)$ representing the remaining off-diagonal contributions.
The state vector $\left\vert \Psi(t) \right\rangle$ and the matrix representation of the
Hamiltonian $H(t)$ can then be written as
\begin{equation}
| \Psi(t) \rangle =\left[
\begin{array}{c}
P(t) \\
Q(t)%
\end{array}%
\right] , H(t)=\left[
\begin{array}{cc}
h(t) & R(t) \\
W(t) & D(t)%
\end{array}\right] ,
\end{equation}%
where the $1\times 1$ matrix $h(t)$ and $(n-1)\times (n-1)$ matrix $D(t)$ are the
self-Hamiltonians in the subspaces of $P(t)$ and $Q(t)$, respectively, while
the off-diagonal contributions $R(t)$ and $W(t)$ denote $1\times (n-1)$ and
$(n-1)\times 1$ matrices, respectively.

Let $p(t)=\exp [-i\int\nolimits_{0}^{t}h(s^{\prime })ds^{\prime }]P(t)$. In
the selected one-dimensional subspace, the projected Schr\"odinger equation (see, e.g., Ref.~(\cite{Jing:16})
will imply that $p(t)$ satisfies
\begin{equation}
\dot{p}(t)=\int\nolimits_{0}^{t}\tilde{g}(t,s)p(s)ds,  \label{Eq9}
\end{equation}%
where $\tilde{g}(t,s)$ is the propagator given by $\tilde{g}(t,s)=-g(t,s)\exp
[-i\int\nolimits_{s}^{t}h(s^{\prime })ds^{\prime }]$, with $g(t,s)=R(t)G(t,s)W(s)$,
$G(t,s)=\Gamma _{\leftarrow }\{\exp [-i\int\nolimits_{s}^{t}D(s^{\prime }){%
ds^{\prime }}]\}$, and $\Gamma _{\leftarrow }$ denoting the time-ordering operator.
Specifically,
\begin{equation}
\dot{p}(t)=-\int\nolimits_{0}^{t}g(t,s)e^{-i\int%
\nolimits_{s}^{t}h(s^{\prime })ds^{\prime }}p(s)ds.  \label{Eq10}
\end{equation}%

\section{Pulse control conditions}
%\section{Pulse control conditions.}

If $\dot{p}(t)=0$, the system
will be kept in the subspace $\left\vert \Psi _{1}(t)\right\rangle
\left\langle \Psi _{1}(t)\right\vert $.
By adding a control Hamiltonian $H_{LEO}(t)$ whose control strength
is large enough, we have that $c(s)$ will dominate the exponential term
for $h(s)$ in Eq.~(\ref{Eq10}). Moreover, a strong control function $c(s)$ will ensure that
$g(t,s)p(s)$ varies slowly
compared with $\exp [-i\int\nolimits_{s}^{t}c(s^{\prime })ds^{\prime }]$. Therefore,
Eq.~(\ref{Eq10}) can be simplified to
\begin{equation}
\int\nolimits_{0}^{\tau }ds\exp \left[- i\int\nolimits_{0}^{s}c(s^{{\prime
}})ds^{{\prime }} \right]=0,  \label{eq3}
\end{equation}%
where $\tau $ corresponds to a single control time interval.

Here as an example we consider two types of periodic control:
rectangular pulses and sine function. First for rectangular pulses, the
control function can be taken as%
\begin{equation}
c(t)=\{%
\begin{array}{c}
I , \\
-I , %
\end{array}%
\begin{array}{l}
n\tau <t<(n+1)\tau , \, \text{for }n \text{ even}, \\
\text{otherwise,}%
\end{array}
\label{eq4}
\end{equation}%
where $I$ represents the control strength and $\tau $ is the running time
for applying a single (either positive or negative) pulse. Notice that $c(t)$ can be realized by a
sequence of fast pulses through an arbitrary physical setup.
Inserting Eq.~(\ref{eq4}) into Eq.~(\ref{eq3}), we obtain~\cite{Paval2016}
\begin{equation}
I\tau =2\pi m, \, \text{for } m=1,2,3,....  \label{eq6}
\end{equation}
If the rectangular control pulses satisfy the above conditions, the
transition from the state $\left\vert \Psi _{1}(t)\right\rangle $ to other
states will be restrained.
For sine function, we have
\begin{equation}
c(t)=I\sin (\omega t).  \label{eq5}
\end{equation}
Eq.~(\ref{eq3}) is equivalent to
\begin{equation}
\int\nolimits_{0}^{\tau }ds\exp [\frac{iI}{\omega }\cos (\omega s)]=0.
\end{equation}
Let $\omega \tau =\pi $. Then, equation above becomes
\begin{equation}
J_{0}(I\tau /\pi )=0,  \label{eq7}
\end{equation}%
where $J_{0}(x)$ is the zero order Bessel function of the first kind.

\section{AEST under LEO control in spin chains}
%\section{AEST under LEO control in quantum spin chains.}

Now we will show that, by choosing an appropriate LEO Hamiltonian, AEST can
be realized with high fidelity. We consider the communication channel as a one-dimensional
XY spin chain, whose Hamiltonian reads
\begin{equation}
H=\sum\limits_{i=1}^{N-1}J_{i,i+1}(\sigma _{i}^{x}\sigma _{i+1}^{x}+\sigma
_{i}^{y}\sigma _{i+1}^{y}),  \label{eq1}
\end{equation}%
where $J_{i,i+1}$ is coupling constant between nearest-neighbor sites and $%
\sigma^k_{i}(k=x,y)$ are Pauli operators acting on the spin at site $i$.
For a uniformly coupled chain, the Hamiltonian $H$ can be represented by $%
H_{UNI}$, with $J_{i,i+1}=J$ is a constant. For simplicity, we take $J=1$.

As an example, we can transfer the state $\left\vert 1\right\rangle $ from one
end of the chain to the other end. Initially all the other spins are in the state $%
\left\vert 0\right\rangle .$ The initial state of the system is then $%
\left\vert 10..0\right\rangle =\left\vert \Psi _{1}\right\rangle .$ Here we
use $\left\vert \Psi _{n}\right\rangle $ to denote that the n\emph{th} spin
is in the up state while all the other spins are in the down state. Define the fidelity
as $F(t)=\sqrt{\left\langle \Psi _{n}\right\vert \rho (t)\left\vert \Psi
_{n}\right\rangle }$, where $\rho (t)$ is the instantaneous density operator of the system. The
fidelity will decrease quickly by increasing the length $N$ of the chain in a
uniform chain~\cite{Bose}.

Our aim is then to provide a robust AEST scheme
through a uniform chain by adding a LEO Hamiltonian.
This will be achieved by considering two types of spin chain Hamiltonians
in Eq.~(\ref{eq1}): (i) PST couplings $J_{i,i+1}=\sqrt{i(N-i)}$ ($H_{PST}$);
(ii) weak couplings $J_{1,2}=J_{N-1,N}=J_{0}$ and $J_{i,i+1}=J$ elsewhere, with $J_{0}\ll J$ ($H_{WC}$).
By directly realizing state transfer through the XY spin chain driven by PST couplings,\ at time $t=k\pi /2$ ($k$ is an integer),
we have that PST occurs, with the state $\left\vert 1\right\rangle $ being perfectly transferred from the first
site to the last~\cite{Christandl}. Similarly, for the spin chain driven by weak couplings, AEST
can also be realized~\cite{Wojcik,Oh, Wang2013}.

In order to implement AEST with LEO control, we take the complete orthonormal basis as
\begin{equation}
\left\vert \Psi _{n}(t)\right\rangle =V(t)\left\vert \Psi _{n}\right\rangle ,
\label{LEObasis}
\end{equation}%
where $V(t)=\exp (-iH_{j}t)$, $(j=PST$ or $WC)$. Clearly $\left\langle \Psi
_{m}(t)\mid \Psi _{n}(t)\right\rangle =\delta _{mn}$ and we will satisfy the
one-to-one relationship between $|\Psi _{n}(t)\rangle$ and $\left\vert \Psi _{n}\right\rangle $.
In what follows, we will consider the AEST in a uniform chain.

\begin{figure}[tbph]
\centering
\includegraphics[width=2.65in]{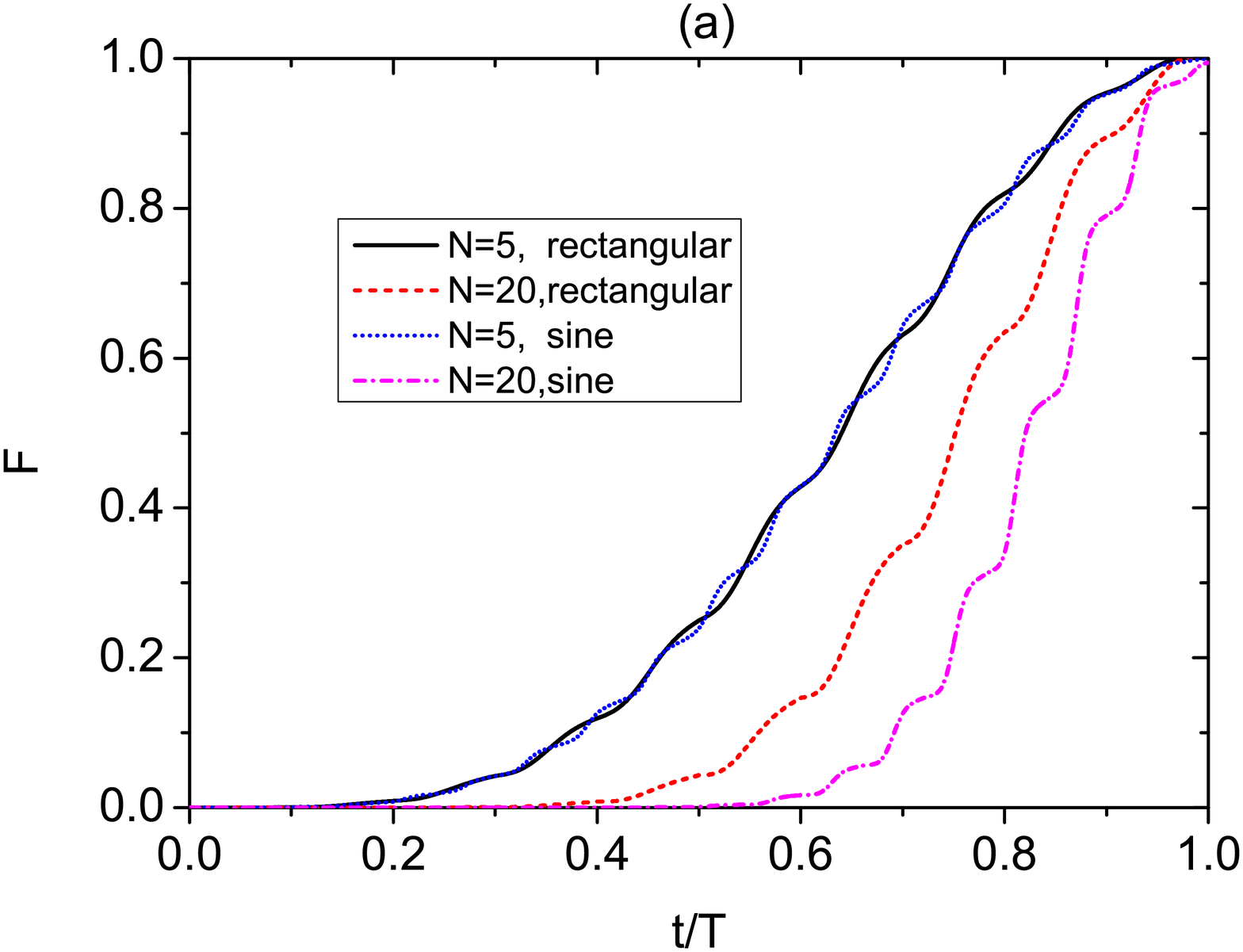} %
\includegraphics[width=2.65in]{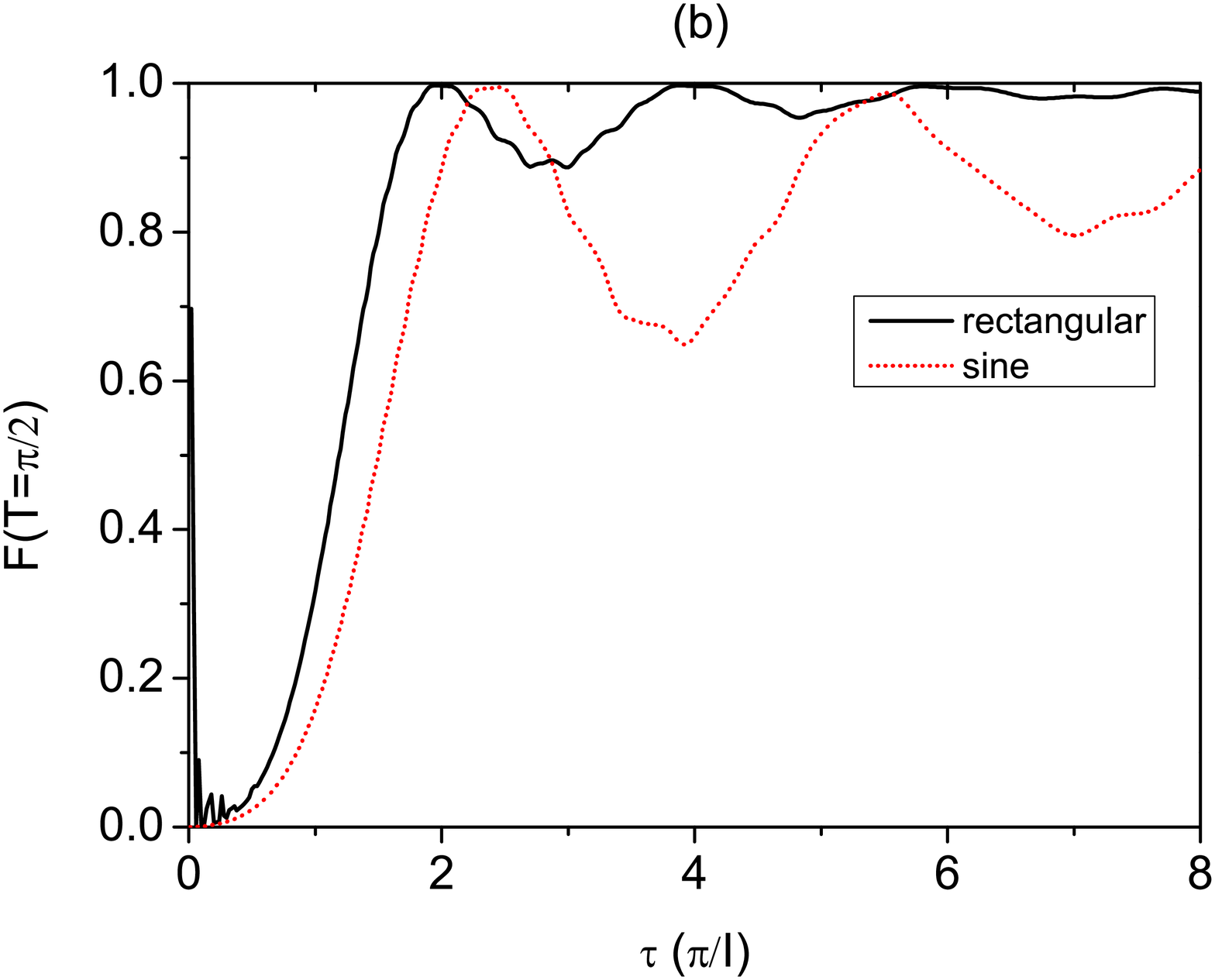}
\caption{Fidelity for AEST under LEO control with PST couplings.
We adopted $T=\protect\pi /2$.
(a) Fidelity $F$ as a function of the normalized time $t/T$
for either rectangular pulses or sine function.
For rectangular pulses, we used $I=40$ and $\protect\tau =\protect\pi /20$.
For the sinusoidal case, we used $I=80$ and $\protect\tau =2.405\protect\pi /80$.
(b) Fidelity $F(T=\pi/2)$ as a function of the control time interval $\protect\tau $ for
either rectangular pulses or sine function. For both cases, we used $I=50$ and $N=10$.
Note that the peak of $F$ occurs at $I\protect\tau =2m\protect\pi ,m=1,2,3,...$ for rectangular pulses
and $I\protect\tau =x\protect\pi $, with $x=2.405, 5.520,...$, for sine function.
The value of $x$ is the zero point of the Bessel function of the first kind $J_{0}(x)$.}
\label{fig1}
\end{figure}

\section{Discussion}

%\section{LEO Hamiltonian generated by PST coupling Hamiltonian.}
First, let us take PST couplings in Eq.~(\ref{LEObasis}), namely, $V(t)=\exp [-iH_{PST}t]$.
We then plot the fidelity $F$ as a function of the normalized time $t/T$ in Fig.~\ref{fig1}(a), where the
density operator is exactly obtained through the solution of the time-dependent Sch\"odinger equation.
For either rectangular pulses or sine function, the total evolution time is taken as $T=\pi /2$.
For rectangular pulses, we take the amplitude as $I=40$ and the time period as $\tau =\pi /20$ according to
Eq.~(\ref{eq6}), while for sine function, we take $I=80$ and $\tau =2.405\pi
/80$, according to Eq.~(\ref{eq7}). We analyze chains with lengths $N=5,20$.
Notice that, from $t=0$ to $t=T$, $F$ increases from zero to nearly unity for both kinds of pulses and
independently of $N$. Then, at time $t=T$, AEST can be realized with high fidelity.

In Fig.~\ref{fig1}(b) we plot $F(t=\pi /2)$ as a function of $\tau $.
The fidelity $F$ initially increases as we increase $\tau$ and then it shows an oscillating
behavior. These two curves have peaks for certain periods $\tau$. The
parameters are set as $T=\pi /2$, $I=50$, and $N=10$.
For rectangular pulses, the peaks correspond to $\tau =(2,4,6,8,...)\pi /I$.
For sine function, they are $\tau=x\pi /I,x=2.405,5.520,8.654,...$.
It is worth noting that the values for $\tau$ satisfy Eqs.~(\ref{eq6}) and (\ref{eq7}),
in agreement with the theoretical prediction by the Feshbach P-Q partitioning technique.

\begin{figure}[tbph]
\centering
\includegraphics[width=2.61in]{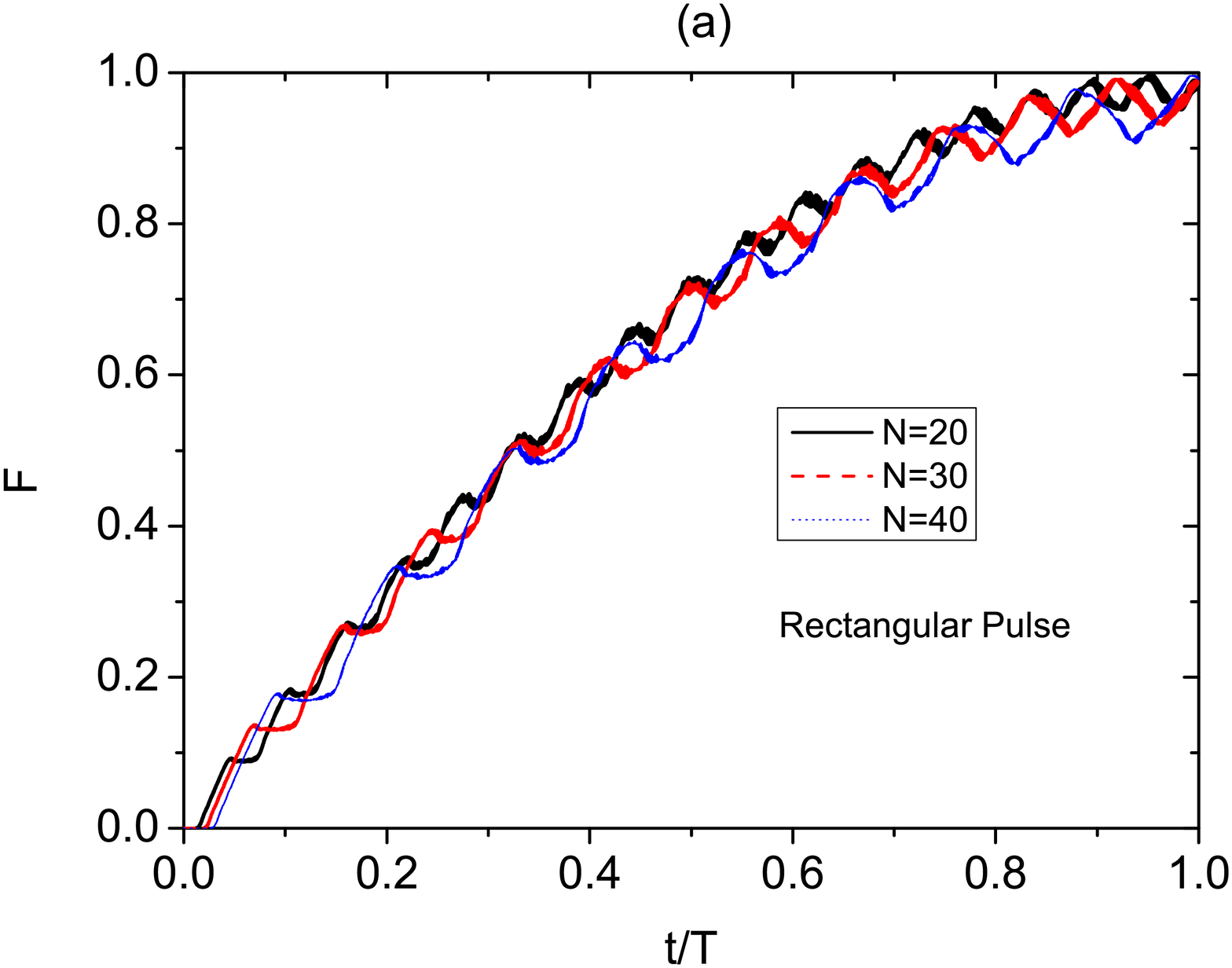}
\includegraphics[width=2.61in]{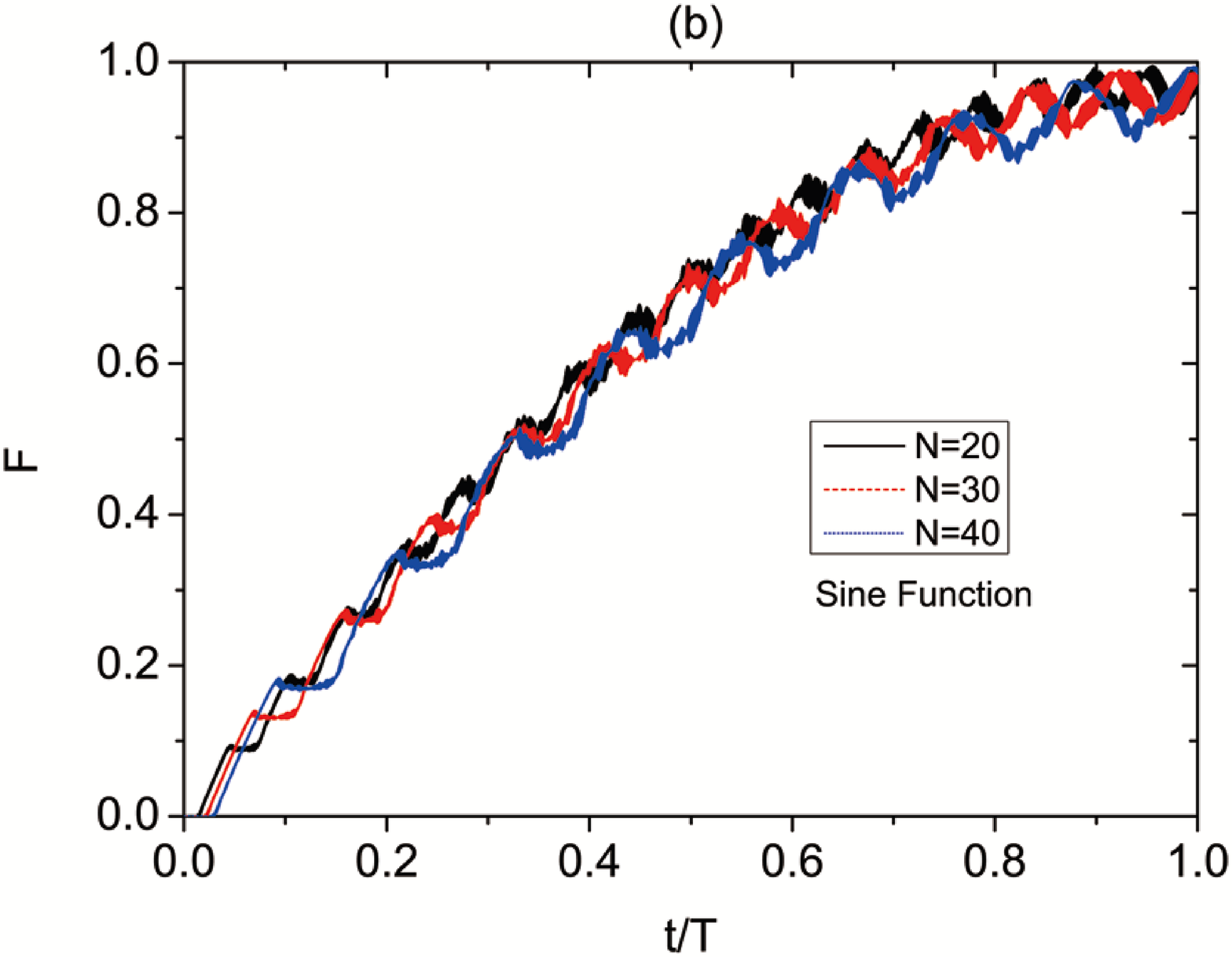}
\caption{
Fidelity for AEST under LEO control with WC couplings.
We adopted $T=210 \pi$.
The fidelity $F$ is plotted as a function of the normalized time $t/T$ for different kinds of pulses and chain sizes $N$.
(a) Rectangular pulses. We used $I=60$ and $\protect\tau =\protect\pi /30$.
(b) Sine function. We used $I=120$ and $\protect\tau =2.405\protect\pi /120$. }
\label{fig2}
\end{figure}
%\section{LEO Hamiltonian generated by WC coupling Hamiltonian.}
Let us now suppose the LEO control is generated by the weak coupling Hamiltonian
$V(t)=\exp[-iH_{WC}t]$. In Figs.~\ref{fig2}(a) and (b), we then plot the fidelity $F$ versus the normalized time $t/T$
for either rectangular pulses or sine function. We consider chains with lengths $N=20,30,40$.
The evolution time is $T=210\pi $.
For both kinds of pulses, AEST ($F=0.999$) can be obtained at time $T$
by effective pulse control. For rectangular pulses, we used $I=60$ and
$\tau =\pi/30$, which satisfies Eq~(\ref{eq6}). For sine function, we used
$I=120$ and $\tau=2.405\pi /I$. This is the first zero value of the Bessel function
of the first kind $J_{0}(x)$. Notice that Fig.~\ref{fig2} again illustrates the effectiveness
of the pulse control scheme, with the only requirement being the adoption of a LEO
Hamiltonian implementing PST or near PST passage.

\begin{figure}[tbph]
\centering
\includegraphics[width=2.61in]{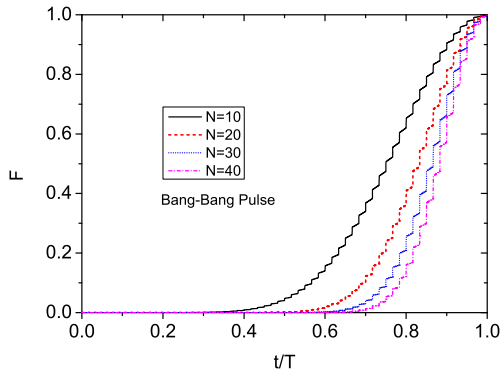}

\caption{Fidelity for AEST under the B-B pulse control with PST couplings for different $N$.
We adopted $T=\pi/2$.
 The pulses are present at $\tau/50$ and otherwise absent in a time interval $\tau =\pi /120$. The integration in $\tau$ satisfies $I\tau=\pi$.}
\label{fig3}
\end{figure}

Until now, the pulses we used are finite and continuous in the above discussion, i.e., at any time the
pulses are present on the chain even though they alternate their directions. This can be regarded as a limit case.
The opposite (second) limit is to utilize a discontinuous and strong pulse sequence, such as B-B kicks (pulses)  $c(t)=\pi \sum_{i}(-1)^{i}\delta(t-\tau_{i})$~\cite{Paval2016}.
As in the first limit case,  again we can use the zero-energy-change pulse. Two consecutive B-B pulses take positive and negative values
respectively. We point out that this case is equivalent to the case that all the negative pulses become
positive~\cite{Paval2016}. In the numerical calculation we use rectangular pulses and LEO Hamiltonian
with PST coupling to simulate the $\delta$ function pulse. Again the total evolution time is taken as $T=\pi/2$.
For a continuous pulse, we take $I=120$ and $\tau=\pi/120$. Now we use $c(t)=50 I$, for $n\tau<t<(n+1/50)\tau$
(odd $n$), $c(t)=-50 I$ for $n\tau<t<(n+1/50)\tau$ (even $n$), and $c(t)=0$, otherwise.
Note that, for the above choice of $c(t)$, the integration still satisfies $\int_{0}^{\tau} c(s)ds=\pi$ in a time interval
$\tau$. Even though $I\gg J$, $c(t)$ oscillates in the range $[-I,I]$.
It is worthy emphasizing that the second limit is experimentally feasible, e.g. using dynamical-decoupling techniques to
moderate the dephasing effects of low-frequency noise on a superconducting qubit~\cite{David}.
In Fig.~\ref{fig3}, we plot the fidelity $F$ versus the normalized time $t/T$ for the simulated B-B pulses.
We consider chains with lengths $N=10,20,30,40$. The results again show that the simulated B-B pulses are effective to realize AEST in a uniform chain.

\section{Conclusions}

 High fidelity quantum state transfer through
spin chains is a powerful tool for performing short-distance quantum
communication in a quantum network. We have
introduced an effective pulse control scheme to realize AEST in a
uniform chain by adding a LEO Hamiltonian to the evolution. The LEO
Hamiltonian can be represented by a sequence of pulses acting on (near) PST
subspaces. By using the Feshbach PQ partitioning technique, we obtained the
conditions over the pulses to allow AEST, which prevent
transitions from the target state to other undesired states.
As an example, we illustrated two kinds of pulses in the XY spin chain: rectangular pulses and sine function.
For both pulses, the relations of the
pulse intensity and its duration is obtained. Numerical analysis explicitly shows
that, once the pulse satisfies the required condition, AEST can
be successfully obtained in a uniform chain setup.

\section{Acknowledgements}

This material is based upon work supported by NSFC (Grant Nos. 11475160,
61575180,11575071) and the Natural Science Foundation of Shandong Province
(Nos. ZR2014AM023, ZR2014AQ026).
M.S.S. is supported by Conselho Nacional de Desenvolvimento Cient\'{\i}fico e Tecnol\'ogico (CNPq-Brazil).
M.S.S. also acknowledges financial support in part by the Coordena\c{c}\~ao de Aperfei\c{c}oamento de
Pessoal de N\'{\i}vel Superior - Brasil (CAPES) (Finance Code 001) and by the Brazilian
National Institute for Science and Technology of Quantum Information [CNPq INCT-IQ (465469/2014-0)].
L.-A.W. is supported by the Basque Country Government (Grant No. IT986-16) and PGC2018-101355- B-I00 (MCIU/AEI/FEDER,UE).

\end{document}